\documentclass[twocolumn,
reprint,
linenumbers,
colorlinks = true,
linkcolor = blue,
urlcolor  = blue,
citecolor = blue,
anchorcolor = blue
]{aastex62}
\usepackage{ulem}
\usepackage{bm,orcidlink}
\usepackage{ulem}
\usepackage{amsmath}

\renewcommand{\sc}{{\rm sc}}
\newcommand{\rms}{{\rm rms}}

\graphicspath{{./}{figures/}}

\renewcommand{\sb}[1]

\begin{document}
\nolinenumbers

\title{On the Heating of the Slow Solar-Wind by Imbalanced Alfv\'en-Wave Turbulence from 0.06 au to 1 au: \emph{Parker Solar Probe} and \emph{Solar Orbiter} observations}


\correspondingauthor{Sofiane Bourouaine}
\email{sbourouaine@fit.edu}

\author[0000-0002-2358-6628]{Sofiane Bourouaine}
\author[0000-0002-8841-6443]{Jean C. Perez}
\affiliation{%
 Department of Aerospace, Physics and Space Sciences, Florida Institute of Technology, Melbourne, Florida, 32901, USA.
}%

\author[0000-0003-4177-3328]{Benjamin D. G. Chandran}
\affiliation{%
 Space Science Center, University of New Hampshire, Durham, NH, 03824, USA.
}

\author[0000-0001-6287-6479]{Vamsee K. Jagarlamudi}
\author[0000-0003-2409-3742]{Nour E. Raouafi}
\affiliation{%
 Johns Hopkins University Applied Physics Laboratory, Laurel, MD 20723, USA.
}
\author[0000-0001-5258-6128]{Jasper S. Halekas}
\affiliation{%
Department of Physics and Astronomy, University of Iowa, Iowa City, IA 52242, USA
}

\begin{abstract}
\nolinenumbers
In this work we analyze plasma and magnetic field data provided by the Parker Solar Probe (\emph{PSP}) and Solar Orbiter (\emph{SO}) missions to investigate the radial evolution of the heating of  Alfv\'enic slow wind (ASW) by imbalanced Alfv\'en-Wave (AW) turbulent fluctuations from 0.06 au to 1 au. in our analysis we focus on slow solar-wind intervals with highly imbalanced and incompressible turbulence (i.e., magnetic compressibility $C_B=\delta B/B\leq 0.25$, plasma compressibility $C_n=\delta n/n\leq 0.25$ and normalized cross-helicity $\sigma_c\geq 0.65$). First, we estimate the AW turbulent dissipation rate from the wave energy equation and find that the radial profile trend is similar to the proton heating rate. Second, we find that the scaling of the empirical AW turbulent dissipation rate $Q_W$ obtained from the wave energy equation matches the scaling from the phenomenological AW turbulent dissipation rate $Q_{\rm CH09}$ (with $Q_{\rm CH09}\simeq 1.55 Q_W$) derived by~\cite{chandran09} based on the model of reflection-driven turbulence. Our results suggest that, as in the fast solar wind, AW turbulence plays a major role in the ion heating that occurs in incompressible slow-wind streams.

\end{abstract}

\keywords{TBA}

\section{Introduction} 

Previous in-situ measurements at heliocentric distances near and above 0.3~au revealed the existence of ubiquitous turbulent fluctuations throughout the interplanetary medium (see, e.g., the review of \cite{bruno13}). Most of these turbulent fluctuations have been found to be Alfv\'enic in nature and to propagate mainly outward from the Sun in the fast solar wind \citep{belcher71,tu95}. Remote-sensing observations have also revealed the presence of Alfv\'en waves in the lower corona with sufficient energy to power the solar wind \citep{depontieu07,mcintosh11}. In addition, recent studies have investigated the heating of the solar wind by AWs turbulence using \emph{PSP} measurements \citep{adhikari21,bandyopadhyay23}.

Generally, the solar wind is classified either as fast or slow wind with a typical speed of $V>500$ km/s or $V\leq 500$ km/s, respectively. 
It is widely believed that Alfv\'en-Wave (AW) turbulence may substantially contribute to the heating and acceleration of the fast solar wind. The large-scale fluctuations may not be able to energize the solar wind directly, but they cascade to smaller-scale fluctuations via nonlinear processes (see, e.g., \cite{verscharen19}). At certain scales (near the ion scale), the turbulent fluctuations start to dissipate and provide thermal energy to the plasma. An important nonlinear cascade process occurs between two counter-propagating AWs~\citep{howes13}, and one of the most likely sources of inward-propagating AWs is non-WKB reflection~\citep{heinemann80,velli93,hollweg07}. Several solar-wind models \citep[e.g.,][]{cranmer07,verdini10,chandran11,usmanov14,vanderholst14} and direct numerical simulations of magnetohydrodynamic (MHD) turbulence 
\citep{perez13,dong14,vanballegooijen16,vanballegooijen17,chandran19,shoda19,perez21b,meyrand23} have
investigated how AW turbulence and wave reflections might heat and accelerate the solar wind.

The fractional variations in the density and magnetic-field strength are often quite small in the fast solar wind, and the turbulent fluctuations in the fast solar wind are often observed to be imbalanced, in the sense that most of the turbulent fluctuations consist of Alfv\'en-wave-like fluctuations propagating away from the Sun in the plasma frame. Although some slow solar-wind streams are more balanced and compressible than the fast wind, other slow solar-wind streams have similar levels of compressibility and imbalance as the fast wind. This latter category of slow wind is  referred to as the Alfv\'enic slow wind (ASW). Numerous intervals of ASW have been observed, for example, by the Parker Solar Probe \cite[\emph{PSP};][]{Raouafi23a}
\citep[see also, ][]{bale19,kasper19,chen21,bourouaine20,bourouaine22}.

Although there is broad agreement that AW turbulence plays a key role in the origin of the fast solar wind, the origin of the slow solar wind is still a highly debated topic \citep[][]{abbo16}. Recently, \cite{Raouafi23b} argued that the solar wind is driven by jetting at the source. The magnetic reconnection at the corona naturally generates Alfv\'en waves that might heat and accelerate the solar wind at higher altitudes.
Previous studies suggested that the solar origin of ASW comes from the boundaries of open coronal fields \citep{damicis15,bale19}, and non-Alfv\'enic slow wind could emanate from coronal streamers at the boundary of the heliospheric current sheet \citep{szabo20,chen21}. 

Most previous studies of solar-wind heating by AW turbulence focused on the fast solar wind \citep[e.g.,][]{cranmer09,chandran11,adhikari21,bandyopadhyay23}.
A notable exception to this was the recent study by \cite{halekas23}, who quantified the radial profiles of the various contributions to the solar-wind energy flux in both the fast solar wind and slow solar wind.

In this paper, we analyze a set of measurements from \emph{PSP} (for heliocentric distance $r=[0.06,0.3]$ au) and Solar Orbiter (\emph{SO}) (for $r=[0.3,1]$ au) to investigate the heating of incompressible slow solar-wind streams by imbalanced AW turbulence. We use the steady-state electron and the proton energy equations to estimate the electron and proton heating rates. For the estimation of the AW turbulent dissipation rate, we use the fluctuation-energy equation in the steady state given in~\citet{perez21b}. Our analysis differs from that of \cite{halekas23} in that we isolate the effects of AW dissipation and plasma heating, rather than quantifying the total plasma energization resulting from the combination of AW dissipation and the work done by AWs.
 In section~\ref{sec:methodology}, we describe our methodology and data analysis for the estimation of the plasma and turbulent parameters. The main results are presented in section~\ref{sec:results}. Finally, in section~\ref{sec:discussion}, we summarize and discuss our findings.

\section{Methodology and data analysis\label{sec:methodology}}

In our analysis, we use plasma and field measurements from \emph{PSP} and \emph{SO}. The data from \emph{SO} (from August 8 2022 to January 31 2023) are used to study the radial profile of the plasma heating rate and the AW turbulent heating rate between 0.3 au and 1 au. We use data from Encounters 2,4,5,6,7,9 and 10 of \emph{PSP} for the analysis between 0.06 au and 0.3 au. These Encounters are chosen due to the availability of plasma density that is estimated through the quasi-thermal-noise (QTN)~\citep{moncuquet20}. 

\subsection{Selection of intervals}

In our analysis, we investigate the radial evolution of the solar-wind plasma as well as the AW turbulent dissipation rate. We divide \emph{SO} and \emph{PSP} data into 9 hours intervals, only intervals corresponding to 9-hours averaged solar-wind speed $V<500$ km/s are initially considered.
Both \emph{SO} and \emph{PSP} measure solar-wind plasma near the ecliptic; therefore, there will be occasions for some intervals with a mixed field polarity. To eliminate those intervals, we only consider intervals in which at least 70\% of the instantaneous $B_r$ values have the same sign, where $B_r$ is the radial component of the magnetic field. The time duration of each interval is chosen to be $T=9$ hours so that the large-scale part of the turbulence can be recovered. For those slow-wind selected intervals, the background (averaged) magnetic-field vector $\bm B_0$ is properly estimated. Then we estimate the turbulence parameters, such as the plasma compressibility, $C_n=\delta n_{\rms}/\langle n\rangle$; the magnetic compressibility, $C_B=\delta B_{\rms}/\langle|B|\rangle$; and the normalized cross helicity, 
\begin{equation}
\sigma_c=\frac{ \langle2\delta {\bm V} \cdot \delta {\bm b}\rangle}{(\langle\delta{V^2}\rangle+\langle\delta{b}^2\rangle)}.
\end{equation}
Here $\delta n_{\rms}$ and $\delta B_{\rms}$ are the root mean square of the fluctuating number density $n=n_e$ (here we assume $n_e\simeq n_p$, where $n_e$ and $n_p$ are the electron and proton density), and the magnetic field strength, respectively. The symbol $\langle\cdots\rangle$ denotes an average over the time period $T$ for each selected interval. The quantities $\delta {\bm V}$ and $\delta {\bm b}$  are the fluctuating bulk velocity and the fluctuating Alfv\'en velocity (i.e., $\delta {\bm b}= \delta {\bm B}/\sqrt{4\pi\rho_0}$, where $\rho_0$ is the proton mass density averaged over the time interval~$T$.) 

These parameters allow us to select intervals of Alfv\'enic slow solar wind (ASW). We consider ASW intervals to be those intervals that are less compressible and where AW turbulence is imbalanced (i.e., by choosing $|\sigma_c|>0.65$, $C_B\leq 0.25$ and $C_n\leq 0.25$). 
\begin{figure}[!t]
    \centering
     \includegraphics{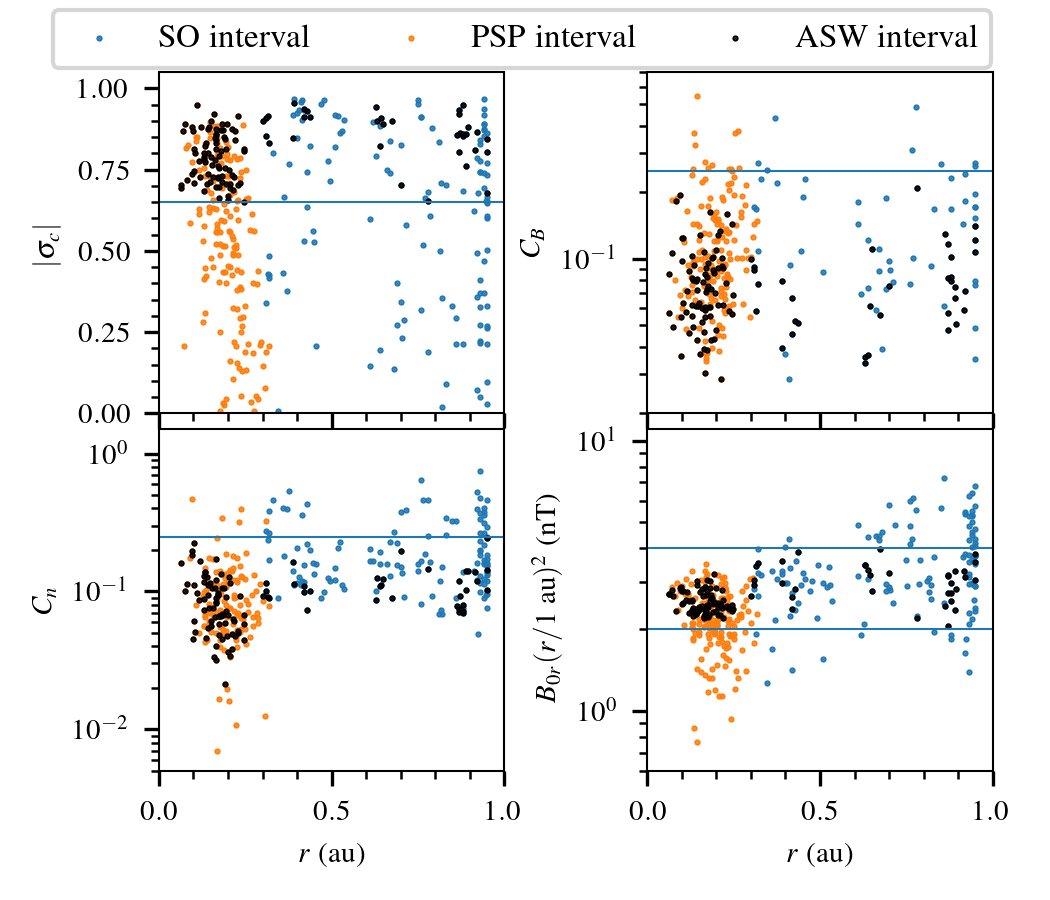}
    \caption{Top left panel: The normalized cross-helicity $|\sigma_c|$. Top right panel: The magnetic compressibility $C_B$. Bottom left panel: The plasma compressibility $C_n$. Bottom right panel:  The quantity $B_{0r} (r/\mbox{~au})^2$ in nT unit that is proportional to the magnetic field flux. The orange (blue) scatter points are the slow-wind \emph{PSP} (\emph{SO}) intervals with more than 70\% of either sun outward or inward magnetic field polarity. The black dots correspond to the selected ASW intervals used for our analysis, i.e., intervals that satisfy; $C_B<0.25$, $C_n<0.25$, $|\sigma_c|>0.65$ and that $B_{0r} (r/1\mbox{~au})^2$ ranges between 2 nT and 4 nT. }
    \label{fig:fig1}
\end{figure} 
In Figure~\ref{fig:fig1}, we show a scatter plot for $\sigma_c$ (upper left panel), $C_B$ (upper right panel), and $C_n$ (lower left panel) corresponding to all 9-hour slow-wind intervals satisfying the condition that more than 70\% of the instantaneous~$B_r$ values within the interval have the same sign. Overall, about 60\% of $|\sigma_c|$ values are higher than 0.65. Since we are interested in slow-wind intervals with AW turbulence, we apply the low-compressibility condition too such that $C_B<0.25$ and $C_n<0.25$. More than 70\% of slow-wind intervals have $C_B<0.25$ and $C_n<0.25$. To study the evolution of the radial profile of heating and turbulent dissipation rates in ASW we apply another condition to the selected intervals, i.e., we only consider those intervals where the quantity $B_{0} (r/1\mbox{~au})^2 \cos\psi=|B_{0r}|(r/1\mbox{~au})^2$ (proportional to the magnetic flux) is between 2 and 4 nT, as indicated by the two horizontal lines in the lower right panel of Figure~\ref{fig:fig1}. Here $\psi$ is the Parker angle that depends on the solar-wind speed and heliocentric distance $r$ as $\tan{\psi}=\Omega r /V$, where $\Omega=2.7\times10^{-6}$ rad s$^{-1}$ is the solar rotation frequency, and $B_{0r}$ is the radial component of the averaged field ${\bm B_{0}}$. Overall, only $\simeq 28 \%$ of the slow-wind intervals seen by \emph{PSP} and SO (during the data period considered) satisfy the criteria of ASW.

\section{Results\label{sec:results}}
\subsection{Radial profiles of plasma and turbulent parameters for ASW.}

\begin{figure}[!h]
 
    \centering
     \includegraphics{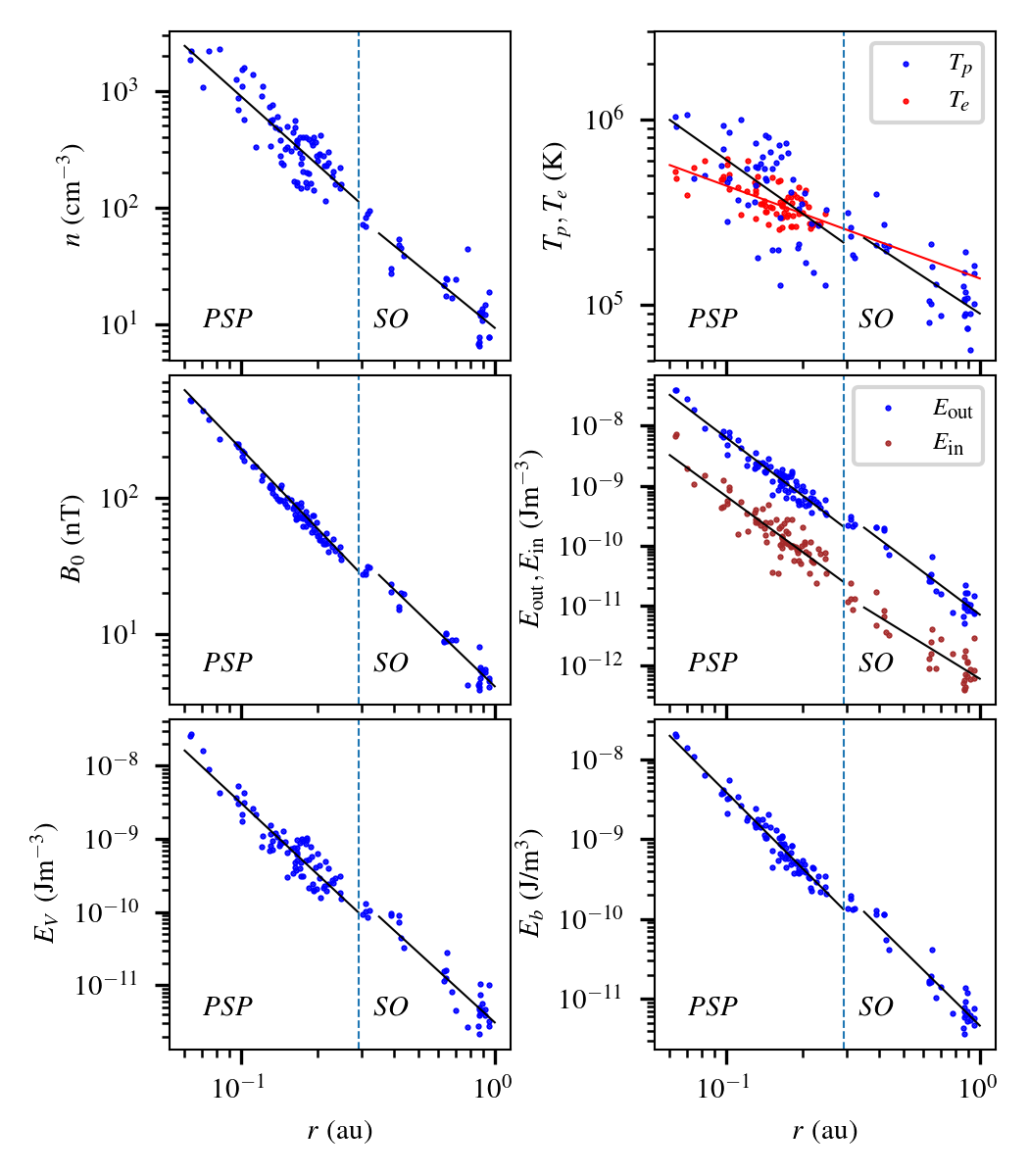}
    \caption{Top panels from left to right: The plasma density $n$ and the proton temperature $T_p$. Middle panels from left to right: The strength of the average magnetic field $B_0$ and the energy densities of the anti-sunward $E_{\rm out}$ sunward $E_{\rm in}$ AWs. Bottom panels from left to right: The energy densities of the kinetic energy density $E_V$ and the magnetic energy density $E_b$ in velocity unit. All plotted as a function of the radial distance $r$. }
    \label{fig:fig2}
\end{figure}

 Using the selected ASW intervals that fulfill all conditions stated above, we can now estimate the radial profiles of the electron density $n$, average proton temperature $T_p$, the strength of the average magnetic field $|\bm B_{0}|$, the energy densities of the anti-sunward (sunward) AWs $E_{\rm out}=\rho_0 \langle\delta Z^2_{\rm out}\rangle/4$ ($E_{\rm in}=\rho_0 \langle\delta Z^2_{\rm in}\rangle/4$), and the energy densities of the fluctuating Alfv\'en velocity $E_b=\rho_0 \langle\delta b^2\rangle/2$ and bulk velocity $E_V=\rho_0 \langle\delta V^2\rangle/2$. Here $\delta {\bm Z}_{\rm out}=\delta {\bm V}-\delta {\bm b}$ ($\delta {\bm Z}_{\rm in}=\delta {\bm V}+\delta {\bm b}$) if the background vector magnetic field is pointing outward from the sun. However, if the background field is pointing toward the sun then $\delta {\bm Z}_{\rm out}=\delta {\bm V}+\delta {\bm b}$ ($\delta {\bm Z}_{\rm in}=\delta {\bm V}-\delta {\bm b}$). All these quantities are plotted in Figure~\ref{fig:fig2}. Each quantity is fit to a power-law function $A_0(r/1\mbox{~au})^{-\alpha}$ for \emph{PSP} and \emph{SO} separately. The fitting parameters are summarized in Table (1) below:

\begin{table}[!h]
\begin{center}

\begin{tabular}{|c| c c | c c|} 

\multicolumn{1}{c|}{} &
\multicolumn{2}{c|}{\emph{PSP}} & \multicolumn{2}{c|}{\emph{SO}} \\ %
 \hline
 $X$ & $A_0$ & $\alpha$ &  $A_0$ & $\alpha$  \\ 
 \hline
 $n$ (cm$^{-3}$) & 10  & -1.95 &  9.35 & -1.8 \\ 
 \hline
 $B_{0}$ (nT) & 2.62  & -1.90 & 4.12  & -1.77 \\
 \hline
 $T_p$ (K) & 6.5e4& -0.95 & 8.9e4 & -0.9 \\
 \hline
 $T_e$ (K) & 1.4e5& -0.5 & - & - \\
 \hline
 $E_b$ (J m$^{-3}$)& 2.6e-12  & -3.15 &  4.6e-12 & -3.12 \\
 \hline
 $E_V$ (J m$^{-3}$) & 1.8e-12  & -3.23&  3.11e-12 & -3.17 \\ 
\hline
$E_{out}$ (J m$^{-3}$) & 3.95e-12 & -3.20&  7.12e-12 & -3.18 \\ 
\hline
 $E_{in}$ (J m$^{-3}$) & 5.52e-13 & -3.10&  6.10e-13 & -2.60 \\ 
\hline
\end{tabular}
\end{center}
\caption{Fitting parameters A, $\alpha$ for each measured quantity $X=A_0 (r/\mbox{1~au})^{-\alpha}$ for Alfv\'enic slow-wind (ASW). }
\label{tab:tab1}
\end{table}
\subsection{Estimation of the proton and electron heating rates in ASW}

To estimate the empirical proton heating rate $Q_p$ we use the fitting functions for the proton temperature $T_p$ and the density $n$. This is to evaluate the derivatives with respect to $r$ in the steady-state proton energy equation (see e.g., \cite{cranmer09})
\begin{equation}
\frac{3}{2}n^{5/3} k_B V_0 \frac{d}{dr}\Biggl(\frac{T_p}{n^{2/3}}\Biggl)
=Q_p+\frac{3}{2}\nu_{ep}nk_B(T_e-T_p),
\label{eq:proton_enr}
\end{equation}
where $k_B$ is Boltzmann constant and $\nu_{ep}$ is the frequency of electron–proton Coulomb collisions. $T_e$ is the electron temperature that is analyzed using only \emph{PSP} data. We found that
\begin{equation}
T_e=1.4\times 10^5 \mbox{~k~} (r/1\mbox{~au})^{\alpha_e},
\end{equation}
with $\alpha_e=-0.5$. This radial profile of $T_e$ will be used to estimate the electron heating rate from 0.06 au to 1 au.
Here, $V_0$ corresponds to the mean solar-wind speed averaged over all ASW intervals, it is $V_0=370$ km/s for ASW \emph{SO} intervals and $V_0=315$ km/s for ASW \emph{PSP} intervals. The collision term in the r.h.s of Eq. (\ref{eq:proton_enr}) is insignificant but still calculated assuming two isotropic Maxwellian distributions for electrons and protons interacting with one another \citep{spitzer53}. The frequency $\nu_{ep}$ is thus estimated from \cite{cranmer09}
\begin{equation}
\nu_{ep}\simeq 8.4\times 10^{-9}\left(\frac{n}{2.5\mbox{~cm}^{-3}}\right)\left(\frac{T_e}{10^5\mbox{~K}}\right)^{-3/2}\mbox{~~s}^{-1}.
\end{equation}

Eq. (\ref{eq:proton_enr}) can be practically simplified by implementing the fitting functions $A_0(r/r_0)^{-\alpha}$ of $n$ and $T_p$ to evaluate the derivatives with respect to $r$. Then, the proton heating rate $Q_p$ is given by
\begin{equation}
Q_p=\frac{3}{2}nk_B \left[V_0 \left(\alpha_p-\frac{2}{3}\alpha_n\right)\frac{T_p}{r}-\nu_{pe}\left(T_e-T_p\right)\right],
\label{eq:Qp}
\end{equation}
where $\alpha_p$ and $\alpha_n$ are the exponents of the power-law fits of $T_p$ and $n$, respectively. Similarly, the radial profile of the electron heating rate $Q_e$ will be estimated by implementing power-law fits of $T_e$, and $n$ into the following steady-state electron energy equation 
\begin{multline}
\frac{3}{2}n^{5/3} k_B V_0 \frac{d}{dr}\Biggl(\frac{T_e}{n^{2/3}}\Biggl)
=Q_e+\frac{3}{2}\nu_{ep}nk_B(T_p-T_e)
\\-\frac{1}{r^2}\frac{\partial}{\partial r}\left(q_{\parallel,e} r^2 \cos{\psi} \right).
\label{eq:electron_enr}
\end{multline}
Here $q_{\parallel,e}$ is the parallel electron heat flux. We estimate $q_{\parallel,e}$ based on the collisionless model~\citep{hollweg74,hollweg76} as
\begin{equation}
q_{\parallel,e}=\frac{3}{2} \alpha_HnV_0k_BT_e,
\label{eq:q_para}
\end{equation}
where $\alpha_H$ is a dimensionless parameter that is only known approximately. For fast wind analysis \cite{chandran11} set $\alpha_H=0.75$. In our analysis of ASW we evaluate $Q_e$ for $\alpha_H=0.5,\mbox{~}0.75 \mbox{~and~0.90} $.
In order to evaluate the derivative of the electron heat flux term in Eq. (\ref{eq:electron_enr}) we use the conservation of the magnetic field flux $B_0 r^2\cos\psi = {\rm constant}$ and the fitting functions of $T_e$, $n$ and $B_0$ that we measured. The electron heating rate is then
\begin{multline}
Q_e=\frac{3}{2}nk_B \left[ V_0 \left(\alpha_e-\frac{2}{3}\alpha_n\right)\frac{T_e}{r}-\nu_{ep}(T_p-T_e)\right]
\\+\frac{q_{\parallel,e}}{r} \cos\psi \left(\alpha_n+\alpha_e-\alpha_B\right)
\label{eq:Qe}
\end{multline}

\begin{figure}[!t]
    \centering
    \includegraphics{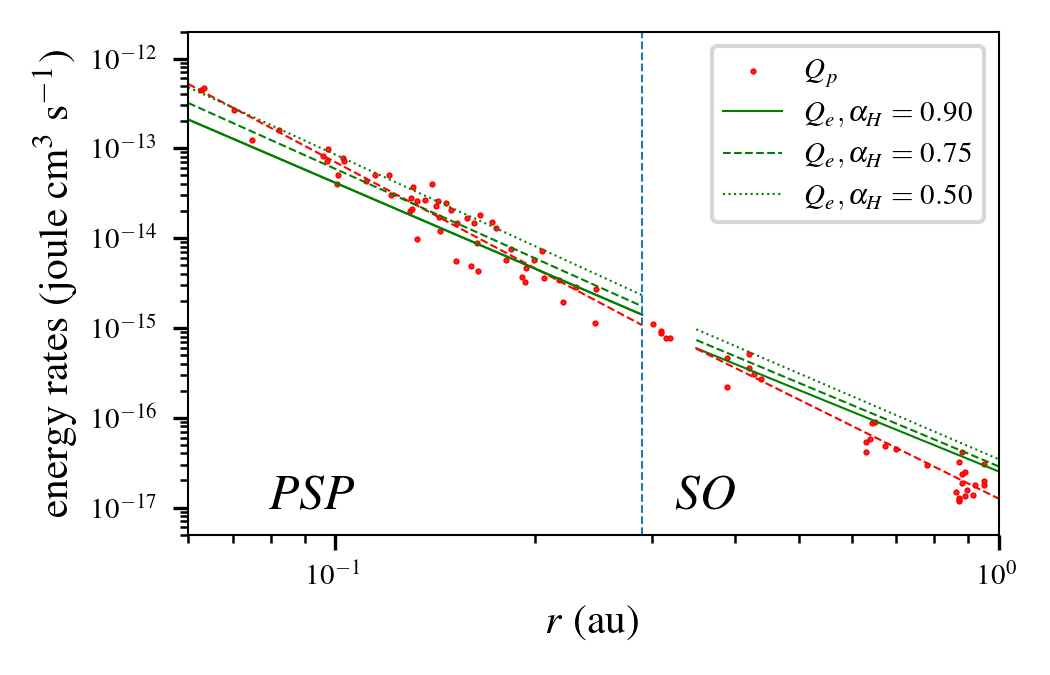}
    \caption{Red dots: Measured proton heating rate $Q_p$ for ASW at $r<0.3$ au ($r>0.3$ au) using \emph{PSP} (\emph{SO}). Green lines: is the estimated electron heating rate $Q_e$ for ASW at $r<0.3$ au ($r>0.3$ au) using $T_e=1.4\cdot 10^5 (r/1\mbox{~au}))^{-0.5}$ (K) for $\alpha_H=0.9$ (solid), $\alpha_H=0.75$ (dashed) and $\alpha_H=0.5$ (dotted). }
    \label{fig:fig3}
\end{figure} 

Figure~\ref{fig:fig3} shows the empirical values of $Q_p$ and $Q_e$ estimated from \emph{PSP} (at $r<0.3$ au) and \emph{SO} (at $r>0.3$ au). We found that the fitting functions of the radial profiles proton rates are  $Q_p=0.83\cdot 10^{-17} (r/1\mbox{~au})^{-3.92}$ J m$^{-3}$ s$^{-1}$  ($Q_p=1.25\cdot 10^{-17} (r/1\mbox{~au})^{-3.66}$ J m$^{-3}$ s$^{-1}$) at $r<0.3$ au ($r>0.3$ au) using \emph{PSP} (\emph{SO}). In Table (2) we show the electron heating rates for different values of $\alpha_H$.

\begin{table}[!h]
\begin{center}

\begin{tabular}{|c| c |c|} 

\multicolumn{1}{c|}{(J m$^{-3}$ s$^{-1}$) } &
\multicolumn{1}{c|}{\emph{PSP} ($r<0.3$ au)}  & \multicolumn{1}{c|}{\emph{SO} ($r>0.3$ au)} \\ %
 \hline

 $Q_e (\alpha_H=0.9$) & $2.75\cdot 10^{-17} \hat{r}^{-3.17}$   &   $2.52\cdot 10^{-17} \hat{r}^{-3.00}$ \\ 
 \hline
  $Q_e (\alpha_H=0.75$) & $2.90\cdot 10^{-17} \hat{r}^{-3.30}$   &   $2.87\cdot 10^{-17} \hat{r}^{-3.10}$ \\ 
 \hline
  $Q_e (\alpha_H=0.5$) & $3.50\cdot 10^{-17} \hat{r}^{-3.38}$   &   $3.44\cdot 10^{-17} \hat{r}^{-3.17}$ \\ 
  \hline
 $Q_p$ & $0.83\cdot 10^{-17} \hat{r}^{-3.92}$   &   $1.25\cdot 10^{-17} \hat{r}^{-3.66}$ \\   
 \hline

\end{tabular}
\label{tab:tab2}
\end{center}
\caption{The heating rates $Q_e$ and $Q_p$ for the Alfv\'enic slow wind (ASW). Here $\hat{r}=(r/1\mbox{~au})$}
\end{table}
The proton heating rate needed for ASW is slightly less than the one for the fast solar wind, $Q^{fast}_p=2.4\cdot 10^{-17} (r/1\mbox{~au})^{-3.8}$ J m$^{-3}$ s$^{-1}$ found by \cite{hellinger11} using \emph{Helios} measurements. 

It is worth-mentioning that formula given in Eq (\ref{eq:q_para}) with $\alpha_H=0.5,\ \mbox{~0.75}$ and $0.9$ guarantees that the conductive heating term is comparable in magnitude to the advective terms in Eq. (\ref{eq:electron_enr}) in our analysis.

\begin{figure}[!t]
    \centering
    \includegraphics{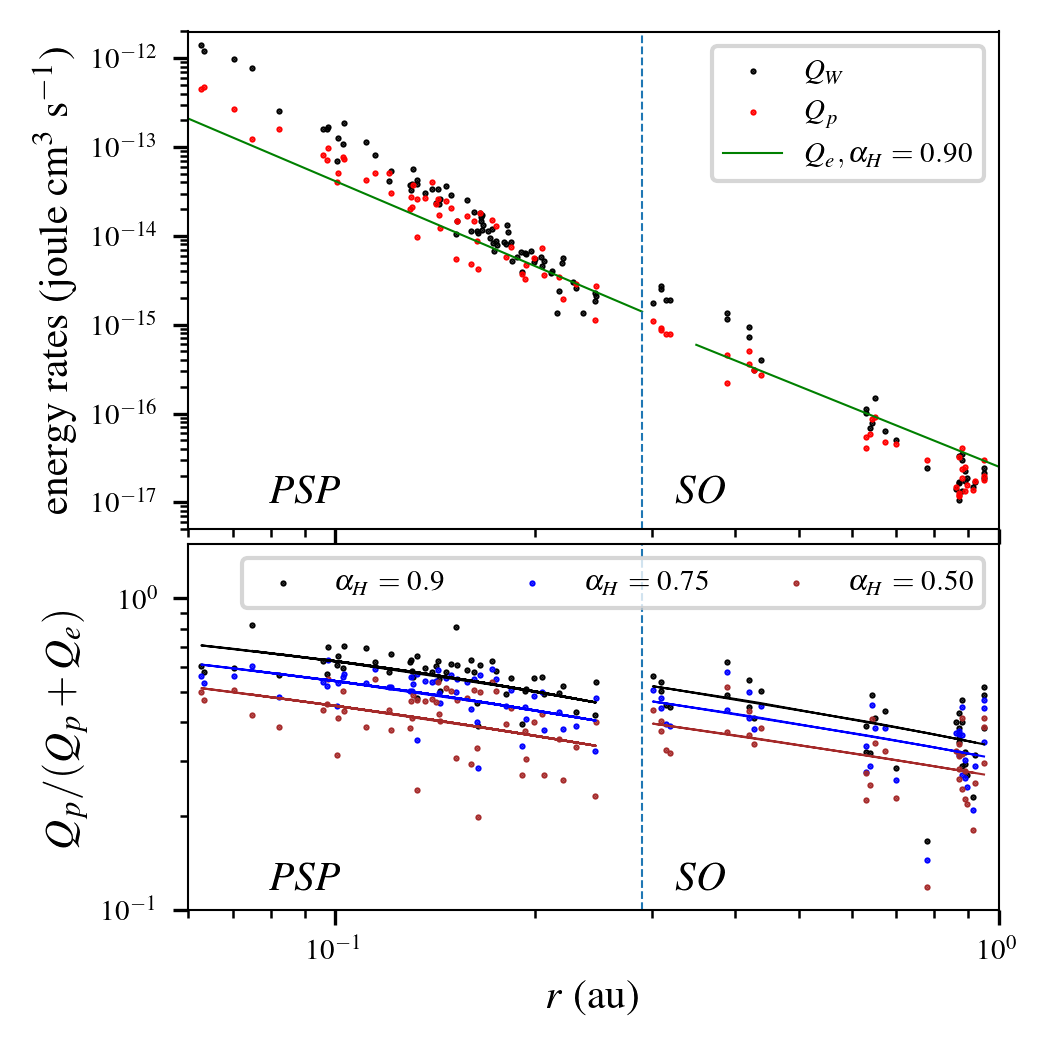}
    \caption{Top panel: Black dots are the estimated turbulent dissipation rate $Q_W$ from Eq. (\ref{eq:Q_W2}) for ASW intervals for $r<0.3$ au ($r>0.3$ au) using \emph{PSP} (\emph{SO}). Red dots represent the proton heating rate $Q_p$ for ASW. Solid green line represents the estimated electron heating rate $Q_e$ when $\alpha_H=0.9$. Bottom panel: proton-to-total heating ratio versus heliocentric distance when $\alpha_H=0.9$, $\alpha_H=0.75$ and $\alpha_H=0.5$, and the solid lines are the same ratio obtained from the power-law fits of $Q_p$ and $Q_e$.}
    \label{fig:fig4}
\end{figure}

\subsection{Estimation of turbulent dissipation rate $Q_W$ in ASW}
To investigate the radial dissipation rate $Q_W$ of the AW (imbalanced turbulence) in the slow-solar wind, we use the steady-state wave-energy equation given in \cite{perez21b} but we account for the Parker spiral field geometry for the flux-tube, thus we have  
\begin{multline}
\frac{d}{r^2dr} \Bigl[r^2 \cos\psi(F^{\rm out}+F^{\rm in})\Bigl]=-Q_w\\
-E_b\frac{d}{r^2dr}(Vr^2 \cos\psi )-\eta V \xi_r
\label{eq:Q_W}
\end{multline}
where $F^{\rm out}=(V+V_A)E_{\rm out}$ ($F^{\rm in}=(V-V_A)E_{\rm in}$) is the flux density of the anti-sunward (sunward) AWs, $\eta=-\cos\psi dB_{0}/(B_{0}dr)$, $\xi_r=E_V-E_b$ is the average residual energy density. Now, we introduce the measured power-law fitting functions of $E_{\rm out}\propto r^{\alpha_{\rm out}}$, $E_{\rm in}\propto r^{\alpha_{\rm in}}$, $|B_{0}|\propto r^{\alpha_B}$ and $n\propto  r^{-\alpha_n}$ to estimate the turbulent heating rate $Q_W$ from Eq. (\ref{eq:Q_W}). The power-law functions and the magnetic-field flux conservation will allow us to simplify the derivatives in Eq. (\ref{eq:Q_W}). Also, by considering that solar wind speed is nearly constant we now can use the following expression of $Q_W$ as 
\begin{multline}
Q_W= -\frac{V_0 \cos\psi}{r}\Bigl[C_{\rm out} (\alpha_{\rm out}-\alpha_B)E_{\rm out}+ 
\\
C_{\rm in}(\alpha_{\rm in}-\alpha_B)E_{\rm in}
- \alpha_B E_v \Bigl],
\label{eq:Q_W2}
\end{multline}
where $C_{\rm out}=1+V_0/V_{A}$, $C_{\rm in}=1-V_0/V_{A}$; and $\alpha_B$, $\alpha_{\rm out}$ and $\alpha_{\rm in}$ are the the power-law fitting exponents of $B_0$, $E_{\rm out}$ and $E_{\rm in}$, respectively.

In Figure~\ref{fig:fig4}, we plot the estimated $Q_W$ (based on Eq. (\ref{eq:Q_W2}) as a function of $r$ using \emph{PSP} (for $r<0.3$ au) and \emph{SO} (for $r>0.3$ au) for ASW intervals. Interestingly enough, the trend of the turbulent dissipation rate $Q_W$ for ASW follows the trend of the proton heating rate $Q_p$ over the entire range of heliocentric distances $r$ we studied. On average, the turbulent heating rate $Q_W$ is about 1.5 times higher than the proton heating rates $Q_p$. The lower panel in Figure~\ref{fig:fig4} shows the ratio $Q_p/(Q_p+Q_e)$ for $\alpha_H=0.9$, $0.75$ and $\alpha_H=0.5$ as a function of $r$.  Overall, It seems that the protons are more heated than electrons in ASW when $\alpha_H\gtrsim 0.9$.  Here we did not consider the collisional Spitzer-H\"arm (SH) electron heat flux as the majority of the observed electron heat flux near-Sun region using {\emph PSP} data lie below the SH limit \citep{halekas21}. Using SH heat flux in our analysis would lead to electron cooling instead of electron heating.


\subsection{Comparison with a phenomenological model of the heating rate in reflection-driven AW turbulence}

\begin{figure}[!t]
    \centering
    \includegraphics{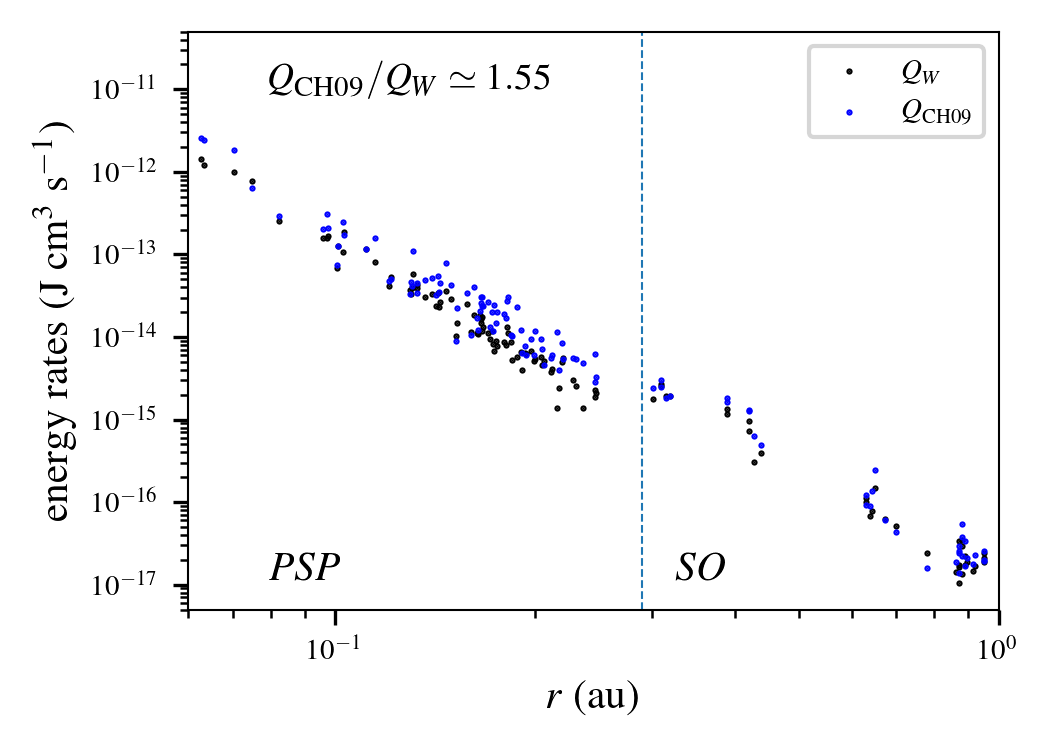}
    \caption{The estimated turbulent dissipation rate, $Q_{\rm CH09}$ from the turbulence reflection model of \cite{chandran09} (blue dots) from Eq.~(\ref{eq:Q_{CH09}}) compared to the turbulent dissipation rate $Q_W$ obtained from Eq. (\ref{eq:Q_W2}) (black dots). All measured using \emph{PSP} for $r<0.3$ au and \emph{SO} at $r>0.3$ au. }
    \label{fig:fig5}
\end{figure}

\cite{dmitruk02} derived a phenomenological turbulent heating rate for reflection-driven turbulence in coronal holes in the absence of background flow. They assumed that there is much more energy in waves propagating away from the Sun than waves propagating towards the Sun, and that the energy cascade timescale of Sunward-propagating waves is comparable to or shorter than their linear wave period.
\cite{chandran09} generalized this model to allow for the background flow of the solar wind, thereby taking into account the work done by the waves on the solar wind.
The essence of these models is to balance the reflection rate of the produced sunward waves against the rate at which these waves cascade and dissipate via interactions with the anti-sunward waves. \citet{chandran09} found that the turbulent heating rate in reflection-driven turbulence is
\begin{equation}
    Q_{\rm CH09} = \frac{1}{4} \left(\frac{ V + v_{\rm A}}{v_{\rm A}}\right) \left|\frac{{\rm d} v_{\rm A}}{{\rm d}r}\right| 
    \rho\left(\delta z^{\rm out}_{\rm rms}\right)^2,
    \label{eq:Q_{CH09}}
\end{equation}
where $V$ is the solar-wind outflow velocity and $v_A$ is Alfv\'en speed. We have estimated the turbulent energy dissipation rate 
$Q_{\rm CH09}$ using the ASW intervals selected for our analysis above. In Figure~\ref{fig:fig5}, we overplot the measured $Q_{\rm CH09}$ together with the estimated $Q_W$ values obtained earlier from the turbulent energy equation (\ref{eq:Q_W2}). This figure shows that the two rates scale in a similar way with~$r$ and are comparable to each other with 
\begin{equation}
    Q_{\rm CH09}\simeq (1.55\pm 0.45) Q_W,
\end{equation}
where the uncertainty $\pm 0.45$ is the propagation error due to the power-law fits. This result provides good observational evidence supporting the reflection-driven turbulence model for turbulent heating in ASW, and is consistent with
a related analysis using \emph{PSP} data by \cite{chen20}.

\section{Conclusion and Discussion\label{sec:discussion}}

In this work we analyzed plasma and magnetic-field data collected by \emph{PSP} and \emph{SO} to study the radial evolution of the plasma heating of the slow solar wind by incompressible Alfv\'enic turbulence. We selected intervals lasting approximately 9 hours in which at least 70$\%$ of the measured magnetic field
has the same polarity -- i.e., the same sign of $B_r$. 
We further restricted our analysis to comparatively incompressible intervals
($C_B< 0.25$ and $C_n< 0.25$) with a high degree of imbalance (i.e., $\sigma_c>0.65$) --- i.e., to ASW.\
We then analyzed those selected intervals to estimate the turbulent heating (dissipation) rate and the plasma (electron and proton) heating rate. We found that the radial profile of the proton heating rate in ASW correlates well with the turbulent heating rate $Q_W$which suggests that Alfv\'enic turbulence plays a major role in the heating of the protons in ASW, as it does in the fast solar wind. We also found that the measured AW turbulent heating rate agrees well with the phenomenological heating rate proposed by \cite{chandran09} for reflection-driven AW turbulence.

Although our results show that AW turbulence can account for much of the heating of ASW between 0.06~au and 1~au, this does not mean that AW turbulence is the dominant mechanism for accelerating the ASW over this range of radii. Indeed, \cite{halekas23} used \emph{PSP} data to examine the radial profiles of the 
various components of the solar-wind energy flux between  $13 R_{\odot}$ and~$100 R_{\odot}$. Because the solar-wind accelerates over this range of radii, the bulk-flow kinetic energy of the solar wind accounts for an increasing fraction of the total energy flux as~$r$ increases. For the slow solar wind, \cite{halekas23} showed that this radial increase is offset primarily by a decrease in the fraction of the energy flux carried by the electron enthalpy flux and heat flux. In other words, it is the electron enthalpy flux and heat flux that account for most of the acceleration of the slow solar wind between $13 R_{\odot}$ and~$100 R_{\odot}$, not AW turbulence.Our results indicate that AW turbulence is likely the dominant heating mechanism, at least in ASW. AW turbulence may heat mainly the protons in ASW through distinct possible heating mechanisms such as resonant-cyclotron heating by high-frequency ion-cyclotron waves  \citep[see e.g., ][]{hollweg02} or by low-frequency AW turbulence \citep[see, e.g., ][]{bourouaine08,chandran13,bourouaine13}. The generation of the high-frequency waves in the solar wind could be triggered by low-frequency turbulence as shown in the numerical simulation work done by \cite{jonathan22}.

\acknowledgments
SB and JCP acknowledge support from NASA grant 80NSSC21K1768. BC acknowledges the support of NASA grants 80NSSC21K1768, NNN06AA01C, and 80NSSC24K0171. VKJ acknowledges support from the Parker Solar Probe mission as part of NASA's Living with a Star (LWS) program under contract NNN06AA01C. 
Parker Solar Probe was designed, built, and is now operated by the Johns Hopkins Applied Physics Laboratory as part of NASA’s Living with a Star (LWS) program (contract NNN06AA01C). Support from the LWS management and technical team has played a critical role in the success of the Parker Solar Probe mission.

\end{document}